\documentclass[twocolumn,lineno,dvipsnames, usenames]{aastex63}
\usepackage[utf8]{inputenc}
\usepackage{amssymb,amsmath,verbatim,mathtools,needspace,enumitem,etoolbox,graphicx,physics,microtype,ulem, breqn}
\usepackage{xspace} 
\usepackage{booktabs}
\usepackage{upgreek} 
\usepackage{acronym}
\usepackage{url}

\newcommand{\LIGOlabMIT}{\affiliation{LIGO Laboratory, Massachusetts Institute of Technology, 185 Albany St, Cambridge, MA 02139, USA}}
\newcommand{\MKI}{\affiliation{Department of Physics and Kavli Institute for Astrophysics and Space Research, Massachusetts Institute of Technology, 77 Massachusetts Ave, Cambridge, MA 02139, USA}}

\newcommand{\pycbc}{\texttt{PyCBC Live}\xspace}

\newcommand{\bilby}{\texttt{bilby}\xspace}

\acrodef{EX}[EX]{\emph{Example}}
\acrodef{bbh}[BBH]{binary black hole}
\acrodef{nsbh}[NSBH]{neutron star--black hole}
\acrodef{bns}[BNS]{binary neutron star}
\acrodef{lvk}[LVK]{LIGO-Virgo-KAGRA Collaboration}
\acrodef{em}[EM]{electromagnetic}
\acrodef{gw}[GW]{gravitational wave}
\acrodef{tess}[TESS]{Transiting Exoplanet Survey Satellite}
\acrodef{fov}[FOV]{field-of-view}
\acrodef{em2}[EM2]{Extended Mission 2}
\acrodef{ffi}[FFI]{full-frame image}
\acrodef{agn}[AGN]{active galactic nucleus}
\acrodef{snr}[S/N]{signal-to-noise ratio}
\acrodef{sed}[SED]{spectral energy distribution}
\acrodefplural{sed}[SEDs]{spectral energy distributions}

\date{\today}

\begin{document}

\title{Searching for Gravitational-Wave Counterparts using the Transiting Exoplanet Survey Satellite}

\correspondingauthor{Geoffrey Mo, Rahul Jayaraman}
 \email{gmo@mit.edu, rjayaram@mit.edu}
\author[0000-0001-6331-112X]{Geoffrey Mo} \MKI \LIGOlabMIT
\author[0000-0002-7778-3117]{Rahul Jayaraman} \MKI
\author{Michael Fausnaugh}\MKI
\author{Erik Katsavounidis}\MKI \LIGOlabMIT
\author{George R. Ricker}\MKI
\author{Roland Vanderspek}\MKI

\begin{abstract}
In 2017, the LIGO and Virgo \ac{gw} detectors, in conjunction with \ac{em} astronomers, observed the first GW multi-messenger astrophysical event, the \ac{bns} merger GW170817.
This marked the beginning of a new era in multi-messenger astrophysics. 
To discover further GW multi-messenger events, we explore the synergies between the \ac{tess} and GW observations triggered by the \ac{lvk} detector network. 
\ac{tess}'s extremely wide field of view of $\sim$2300 deg$^2$ means that it could overlap with large swaths of \ac{gw} localizations, which can often span hundreds of deg$^2$ or more.
In this work, we use a recently developed transient detection pipeline to search TESS data collected during the LVK's third observing run, O3, for any \ac{em} counterparts. 
We find no obvious counterparts brighter than about 17th magnitude in the \ac{tess} bandpass. 
Additionally, we present end-to-end simulations of \ac{bns} mergers, including their detection in \acp{gw} and simulations of light curves, to identify \ac{tess}'s kilonova discovery potential for the \ac{lvk}'s next observing run (O4).
In the most optimistic case, \ac{tess} will observe up to one \ac{gw}-found \ac{bns} merger counterpart per year.
However, \ac{tess} may also find up to five kilonovae which did not trigger the \ac{lvk} network, emphasizing that \ac{em}-triggered \ac{gw} searches may play a key role in future kilonova detections.
We also discuss how \ac{tess} can help place limits on \ac{em} emission from binary black hole mergers, and rapidly exclude large sky areas for poorly localized \ac{gw} events.
\end{abstract}


\section{Introduction}

\begin{figure*}
\centering
	\includegraphics[width=1\linewidth]{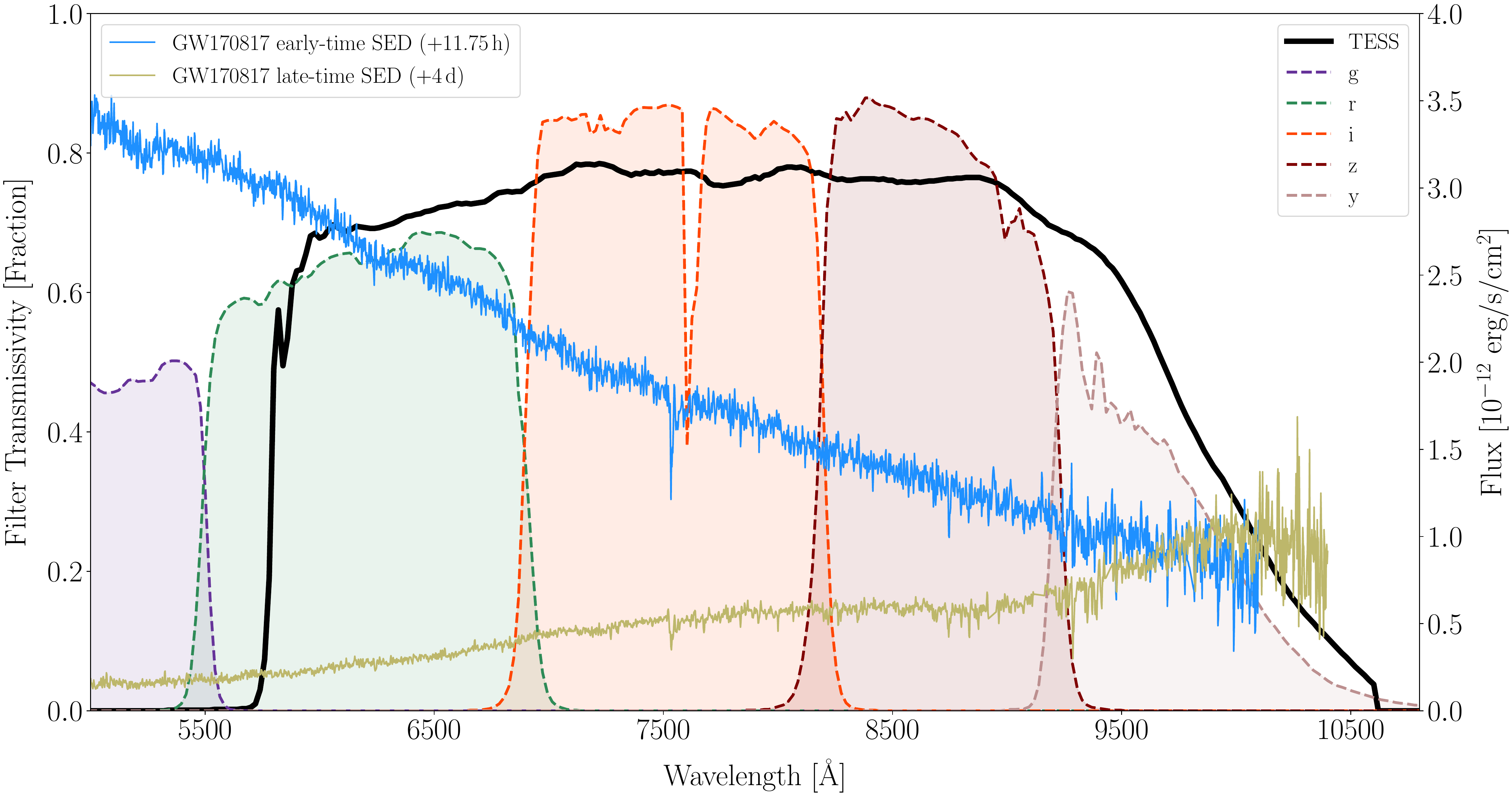}
\caption{The \ac{tess} band (black line) compared to the optical and near-infrared bands of Pan-STARRS \citep{panstarrs-photometric-system} (dotted lines). 
These have been plotted alongside the early-time (light blue) and late-time (tan) \acfp{sed} of the optical counterpart to GW170817 \citep{shappee-sed}.
The \ac{sed} of the kilonova becomes redder as time passes, showing that \ac{tess}'s bandpass is uniquely positioned to continuously observe the decay of the afterglow over a timescale of days. 
}
\label{fig:bands}
\end{figure*}

\label{sec:intro} 
In 2017 August, the LIGO \citep{TheLIGOScientific:2014jea} and Virgo \citep{TheVirgo:2014hva} gravitational-wave (GW) observatories detected the \acf{bns} merger GW170817 \citep{TheLIGOScientific:2017qsa}, which was coincident with the Fermi Gamma Ray Space Telescope's detection of GRB170817A \citep{Goldstein:2017mmi}.
This event ushered in the age of \ac{gw} multi-messenger astronomy, with detections of the resulting kilonova and afterglow in the X-ray, UV, optical, infrared, and radio \citep[and references therein]{GBM:2017lvd, Ciolfi:2020huo}.
The \ac{lvk}'s last observing run, O3, brought the total number of detected \ac{gw} events to 90.
Only one of the events detected during O3 had a potential \ac{em} counterpart \citep{2020PhRvL.124y1102G}; no \ac{em} counterparts were detected for any other events \citep{LIGOScientific:2021djp}.

The next observing run for the \ac{lvk}, O4, is expected to begin in 2023.
With significantly upgraded sensitivity and an expanded detector network (with the addition of KAGRA \citep{KAGRA:2020agh, Aso:2013eba, Somiya:2011np}), O4 promises to bring many more \ac{gw} detections.
\ac{em} counterparts will be searched for across the spectrum by many facilities, including but not limited to optical surveys such as the Zwicky Transient Facility (ZTF; \citealt{Anand:2020eyg}), the Asteroid Terrestrial-impact Last Alert System (ATLAS; \citealt{Tonry2018abc}), and the All-Sky Automated Search for Supernovae (ASAS-SN; \citealt{deJaeger:2021tcq}); targeted infrared searches such as the Wide-Field Infrared Transient Explorer (WINTER; \citealt{Frostig:2021vkt}); wide-field radio follow-up from the Murchison Wide-Field Array (MWA; \citealt{kaplan_murphy_rowlinson_croft_wayth_trott_2016}) and other observatories (see, e.g, \citealt{dobie2019optimised, callister2019first}); as well as rapid X-ray and gamma-ray follow-up from the Neil Gehrels Swift Observatory and the Fermi Space Telescope \citep{oates2021swift, tohuvavohu2020gamma, fletcher2021gamma}.

In this paper, we suggest that another space-based observatory, the \acf{tess}, is poised to play a useful role in \ac{em} follow-up of \ac{gw} detections.
\ac{tess} was launched in 2018 April and has been conducting an all-sky survey since 2018 July, with the primary goal of finding transiting exoplanets around M-dwarfs \citep{tess:ricker}. 
\ac{tess}'s four cameras cover $\sim$2300 deg$^2$ on the sky, observing the same region (a ``sector'') continuously for $\sim 27$ days.
As of 2023 February, it has observed $\sim$ sixty sectors, covering approximately 90\% of the entire sky.
In the context of \ac{gw} follow-up, we highlight \ac{tess}'s enormous \ac{fov} and extensive periods of uninterrupted observation.
The \ac{tess} bandpass spans from 600-1000 nm; see Fig.~\ref{fig:bands} for a visual representation of the \ac{tess} band compared to the set of passbands used in the Panoramic Survey Telescope and Rapid Response System (Pan-STARRS; \citealt{Chambers:2016jzn}). 
\ac{tess}'s broad wavelength coverage spans nearly the entirety of the Pan-STARRS $r$, $i$, $z$, and $y$ passbands; we also show the evolution of the \ac{sed} of GW170817 \citep{shappee-sed}.
The \ac{sed} reddens over a timescale of days, but remains in \ac{tess}'s passband throughout.

\ac{tess}'s Prime Mission (from 2018 July to 2020 July) was successful, with over 2\,000 ``Objects of Interest'' (i.e., planet candidates) detected \citep{guerrero-toi}.\footnote{Since then, the number of \ac{tess} Objects of Interest has ballooned to over 5000 (\url{https://tess.mit.edu/publications/}).}
Data from \ac{tess}, in conjunction with recently-developed software, including the {\tt TESS-reduce} difference imaging pipeline \citep{2021arXiv211115006R}, have also contributed significantly to transient science. 
Key scientific results include population-level studies of Type Ia supernovae \citep{faus-ia-sne}, constraints on the progenitor properties of the double-peaked Type IIb supernova SN2021zby \citep{2022arXiv221103811W}, characterization of the optical afterglow of the gamma-ray burst GRB191016A \citep{2021ApJ...911...43S}, analyses of early-time light curves of tidal disruption events \citep{Holoien:2019zry} and core-collapse supernovae \citep{2021MNRAS.500.5639V}, and identification of periodic \ac{agn} outbursts \citep{asassn-14ko}. 

\ac{tess}'s \ac{em2}, which began in 2022 September, brought with it configuration changes that will prove beneficial to the rapid detection and follow-up of transients.
The two main changes in \ac{em2} are the reduced integration time for each \ac{ffi} (200\,s) and the increased frequency of data downlinks to Earth (weekly instead of biweekly). 
Additionally, during \ac{em2}, \ac{ffi}s will be more frequently released after they have been processed using the TESS Image CAlibrator ({\tt tica}; \citealt{fausnaugh:tica}). 
These factors, in conjunction with our pipeline to identify transients in \ac{tess} \ac{ffi}s (described briefly in Section \ref{sec:trans-id}), will allow for the rapid identification and follow-up of transient candidates, including kilonovae.

In Sec.~\ref{sec:o3}, we present our results from a search of \ac{tess} \ac{ffi} data concurrent with the \ac{lvk}'s O3 to establish preliminary limits on any \ac{em} counterparts to \ac{bns}, \ac{nsbh}, and \ac{bbh} mergers whose \ac{gw} localizations coincided with \ac{tess}'s \ac{fov}.
We then discuss in Sec.~\ref{sec:o4sim} the results of a simulation for the \ac{lvk}'s next observing run, O4, and \ac{tess}'s prospects for observing kilonovae.
We estimate the number of kilonovae detectable in \ac{tess} via both \ac{gw}-triggered searches of the data and blind searches of the \ac{tess} data. 
Finally, in Sec.~\ref{sec:discussion}, we discuss the implications of our results, additional applications of \ac{tess} for \ac{gw} follow-up, and the niche that TESS will occupy in the growing field of multi-messenger astronomy.

\section{TESS observations of O3 events} 
\label{sec:o3}
The \ac{lvk}'s third observing run (O3), which spanned from 2019 April to 2020 March, resulted in a combined 75 \ac{gw} events.
Of these, 39 were released as part of the 2nd Gravitational-wave Transient Catalog (GWTC-2; \citealt{Abbott:2020niy}), and 35 were part of GWTC-3 \citep{LIGOScientific:2021djp}. 
One other event, GW200105\_162426---a marginal \ac{nsbh} candidate---was released separately \citep{LIGOScientific:2021qlt}. 
The events were also released on the Gravitational Wave Open Science Center (GWOSC; \citealt{LIGOScientific:2019lzm}).
69 detections from O3 were confident \ac{bbh} mergers, three were consistent with \ac{nsbh} mergers, and one was a \ac{bns} merger.
Two events had masses which fell in the ``lower mass gap'' \citep{Shao:2022qws}, indicating that they could either be \ac{nsbh} or \ac{bbh} mergers.

Significant \ac{em} follow-up campaigns ensued from the publicly released \ac{gw} events during O3. 
In particular, the \ac{bns} and \ac{nsbh} mergers mentioned above attracted considerable attention from optical and near-infrared observers.
A slew of observatories searched for potential counterparts to these mergers, including ZTF \citep{Coughlin:2020fwx, Anand:2020eyg, Graham:2022xxu}, SAGUARO \citep{Paterson:2020mmd, Lundquist:2019cty}, SkyMapper \citep{Chang:2021zdi}, ASAS-SN \citep{deJaeger:2021tcq}, DECam \citep{Anand:2020mys, Andreoni:2019kqi}, GRANDMA \citep{Antier:2020nuy}, GOTO \citep{Gompertz:2020cur}, GECKO \citep{Kim:2021hhl}, and J-GEM \citep{J-GEM:2021aem}. 
See Appendix A of \cite{LIGOScientific:2021djp} and references therein for details and information on observations in other \ac{em} domains.
Besides the \ac{agn} flare ZTF19abanrhr which \citeauthor{2020PhRvL.124y1102G} theorize to have resulted from an \ac{agn} accretion disk disrupted by the \ac{bbh} merger GW190521 \citep{LIGOScientific:2020iuh}, no \ac{em} counterparts were found by any search.

O3 temporally overlapped with \ac{tess} sectors 10 through 23 (the first two \ac{lvk} observing runs, O1 and O2, occurred before \ac{tess}'s launch).
Sectors 10--13 were located in the southern ecliptic hemisphere, while Sectors 14--23 were located in the North\footnote{
Information about \ac{tess} pointings can be found at \url{https://tess.mit.edu/observations/}. 
Pointings for Sectors 14--16 were adjusted from the typical +54$^{\circ}$ ecliptic latitude to +85$^{\circ}$ to avoid excessive contamination from scattered light from the Earth and Moon in cameras 1 and 2.}.
Each sector was observed for two orbits of 13--14 days each, with a day-long gap during the data downlinks between each orbit. 
During these sectors, \ac{tess} captured \ac{ffi}s of its entire \ac{fov} with an integration time of 30 minutes. 
An estimate for the 3-$\sigma$ limiting magnitude of a single 30\,min \ac{ffi}, for sectors with low backgrounds, is 19.11 in the \ac{tess} band (see, e.g., \citealt{2019GCN.25982....1F}). 
By stacking these \acp{ffi} on timescales of 8 hours or longer, we can probe fainter magnitudes, down to 21--22; a more thorough discussion of such a technique is presented in \citet{rice-stack-limits}.

In this section, we conduct a \ac{gw}-triggered search of the \ac{tess} \ac{ffi}s for each event in GWTC-3 to search for optical and near-infrared counterparts to \ac{bns}, \ac{nsbh}, and \ac{bbh} mergers.
Searching for counterparts to such events will allow us to establish limits on the types of emission we may expect to observe from future events during O4. 
For instance, while \ac{bns} mergers at sufficiently close distances (such as GW170817) are expected to produce detectable kilonovae, we may also observe an electromagnetic counterpart for certain \ac{nsbh} mergers with favorable masses and spins \citep{Zhu:2021ysz, Biscoveanu:2022iue, foucart2020brief}.
On the other hand, \ac{em} emission from \ac{bbh} events is usually not expected \citep{DES:2018uhh}. 

Our search for potential \ac{em} counterparts in \ac{tess} \ac{ffi}s consists of two parts. 
First, we find overlaps between a \ac{gw} probability skymap and the concurrent \ac{tess} sector, as well as the \ac{tess} sectors spanning until 40 days post-\ac{gw} trigger. 
Then, for skymaps in which $>1\%$ of the overall probability is enclosed in the \ac{tess} field of view for either a simultaneous or subsequent sector, we search through \ac{tess} \ac{ffi}s for candidate transient events using our pipeline (Jayaraman et al. in prep). 
Here, we provide a high-level overview of this pipeline; full details are beyond the scope of this work and will be discussed in a separate paper.

\begin{figure*}
    \centering
    \includegraphics[width=\columnwidth, clip=True, trim={0 0 0 0.7cm}]{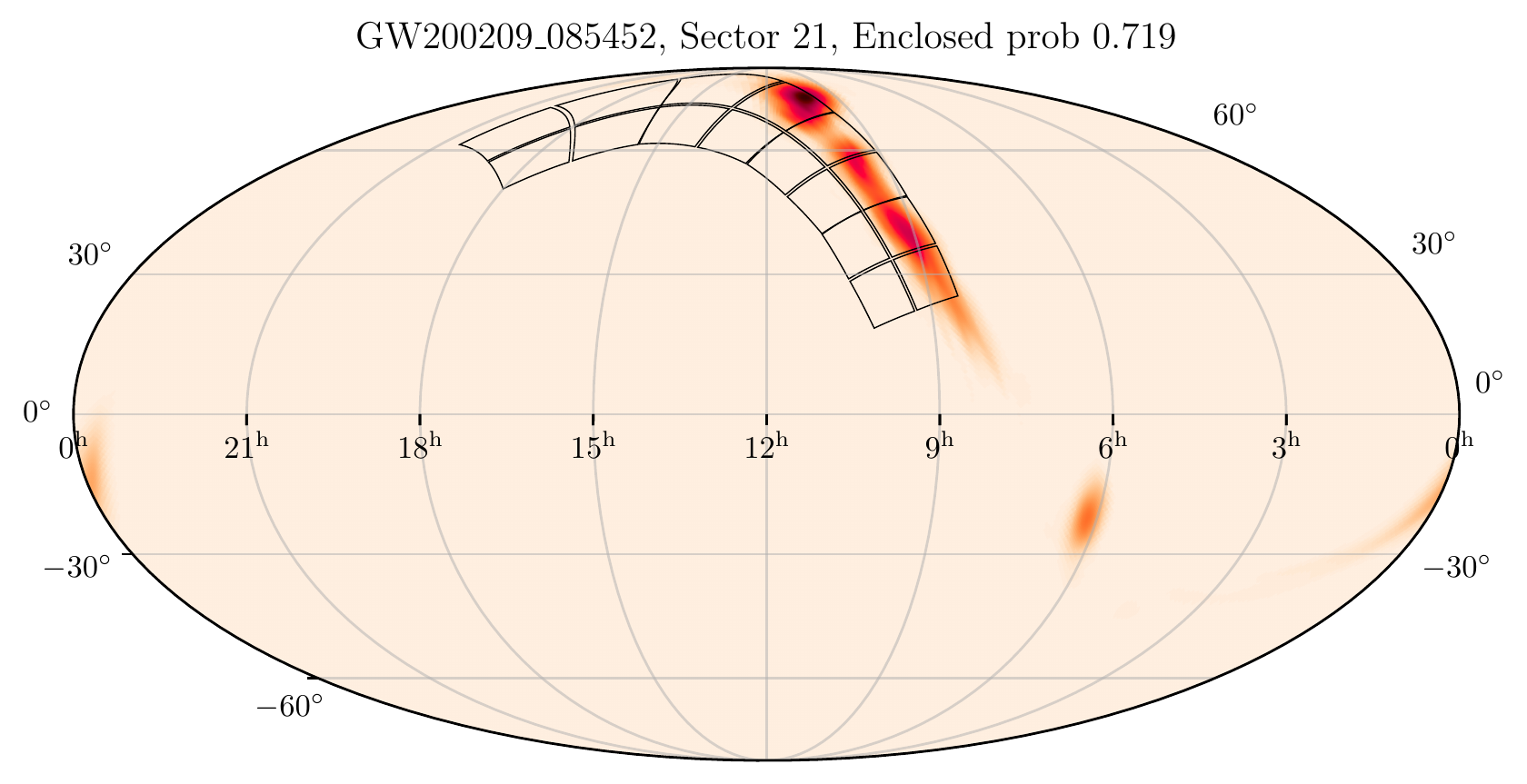}
    \includegraphics[width=\columnwidth, clip=True, trim={0 0 0 0.7cm}]{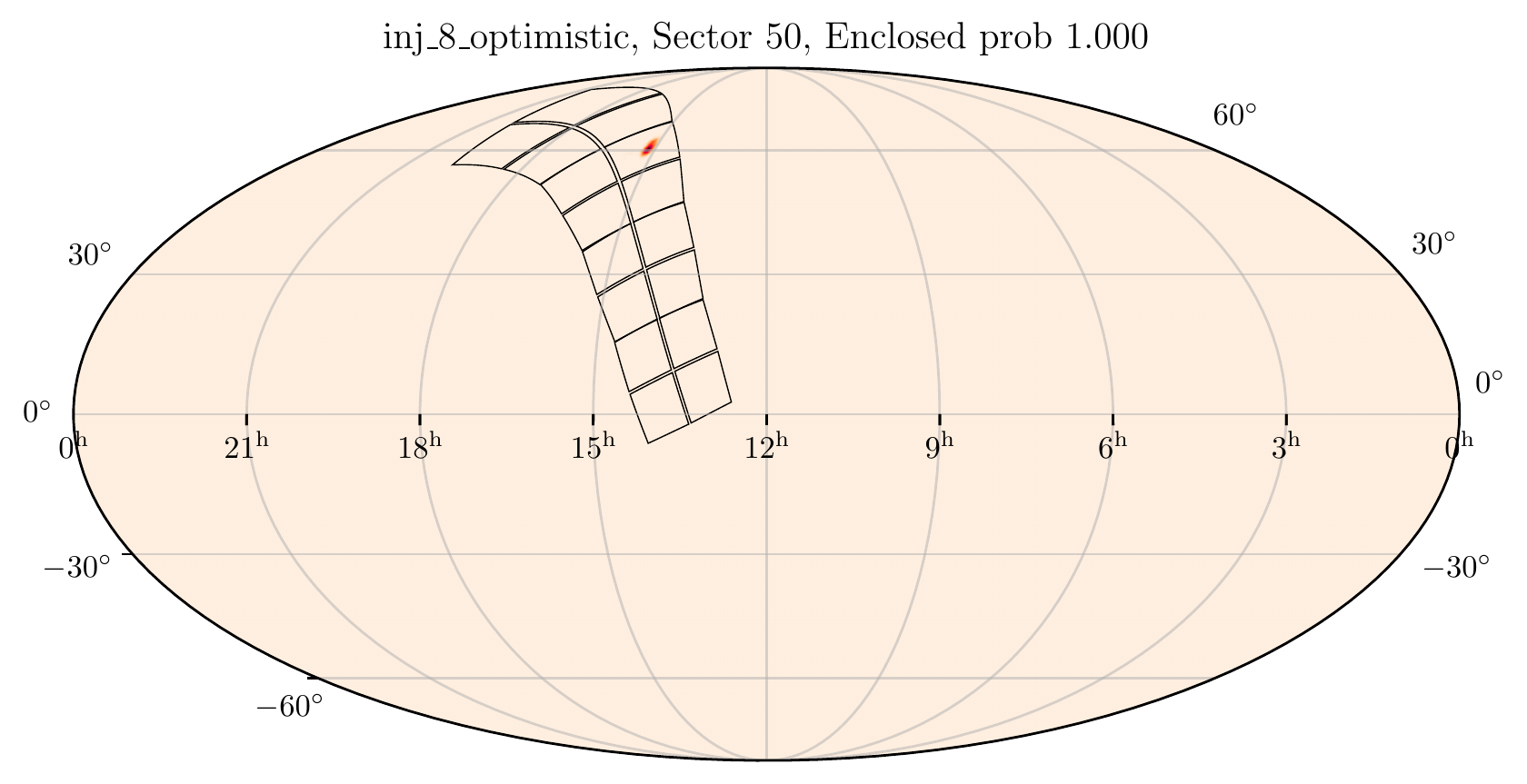}
    \caption{Examples of \ac{tess} coverage of \ac{gw} skymaps, plotted in celestial coordinates, for two individual gravitational-wave events.
    \emph{Left:} The best \ac{tess}-covered event in O3, the \ac{bbh} merger GW200209\_085452, which was concurrent with \ac{tess} sector 21. 
    The enclosed probability is 0.719. 
    No \ac{em} counterparts were found in a search of \ac{tess} data. 
    \emph{Right:} An example simulated \ac{bns} event for O4, where the entire probability is enclosed in the concurrent \ac{tess} sector 50.
    This event's associated simulated light curve peaks at 19.4 mag in the \ac{tess} band, which is easily detectable in \ac{tess} with an 8-hour stack of \acp{ffi} (this 8-hour stack pushes the 3-$\sigma$ detection limit to $\sim$21).
    For an event such as this, the entire \ac{gw} localization region would be observed in the \ac{tess} \ac{fov}, meaning that \ac{tess} would be able to supply not only a precise localization, but also photometry showing the full rise and evolution of the kilonova light curve.
    }
    \label{fig:example_skymap}
\end{figure*}

\subsection{Skymap overlap calculation}
\label{subsec:overlap}
To assess the utility of \ac{tess} in identifying a potential \ac{em} counterpart for a given \ac{gw} event, we developed a tool that constructs and calculates the overlap between a given \ac{tess} sector and a \ac{gw} probability skymap.
For each of the 16 \ac{tess} CCDs, the tool constructs a polygon, accounting for the gaps between the cameras and CCDs.
To translate from detector coordinates to sky coordinates, we use the World Coordinate System (WCS) data in the \ac{ffi} file headers for existing data \citep{2002A&A...395.1061G, 2002A&A...395.1077C}.
For future sectors, we use models from planned pointings (presented in \citealt{kunimoto-pointings}, and also available on the \ac{tess} website).
Each pixel in the \ac{gw} probability skymap is checked to determine if it falls into a \ac{tess} CCD's \ac{fov}, and the total probability enclosed in \ac{tess} is then integrated.
See Fig.~\ref{fig:example_skymap} for two examples of the overlap between a \ac{gw} event and the \ac{tess} \ac{fov}: one from GW200209\_085452, a \ac{bbh} event from O3, and another from a \ac{bns} simulated for O4.

Often, the selected \ac{tess} sector will be the one concurrent with the \ac{gw} observation; however, overlap integrals can also be calculated for any other sector.
For example, while \ac{tess}'s duty cycle is extremely high (at $\sim$90\%), \ac{tess} undergoes a data downlink or is otherwise observationally disrupted approximately 10\% of the time. 
Telemetry interruptions are predictable and can be accounted for.
Sometimes, however, \ac{tess} data collected during nominal operations can be unusable. 
A fraction of \acp{ffi} from every sector (ranging from $\lesssim 10$\% to $\gtrsim 40$\%) suffer from elevated backgrounds due to scattered light from the Earth and the Moon when they are above the spacecraft sunshield.
Scattered light levels in each sector can be qualitatively predicted based on properties of the \ac{tess} orbit, and we can adjust our search strategy accordingly. 
The lowest sky backgrounds occur in sectors that occur between December and March.
All relevant information regarding the \ac{tess} data (formats, disruptions, scattered light, etc.) is documented in the corresponding Data Release Notes.\footnote{For example, see the Cycle 3 DRN summaries: \url{https://heasarc.gsfc.nasa.gov/docs/tess/cycle3_drn.html}}

For events in portions of the data affected by instrument issues or suffering from significant amounts of scattered light, we can calculate the overlap with the next available sector to identify any afterglows or other similar emission.
Another reason to calculate the overlap with subsequent sector(s) is that some models of \ac{em} emission from binary mergers predict somewhat delayed emission; for instance, Equation~3 in \citet{2020PhRvL.124y1102G} suggests that \ac{em} emission from \ac{bbh} mergers can occur tens of days after the coalescence.
Thus, if a \ac{gw} trigger occurs at the end of a sector, it would be prudent to search for evidence of an \ac{em} counterpart throughout the subsequent one to two sector(s), a timespan of $\sim$ 50 days.
While we understand that these timescales remain poorly constrained, we adopt this strategy to serve as a proof-of-concept in our initial search of archival \ac{tess} data. 

\subsection{Searches in Archival Data}
\begin{figure*}[ht!]
    \centering
    \centerline{\includegraphics[width=\textwidth]{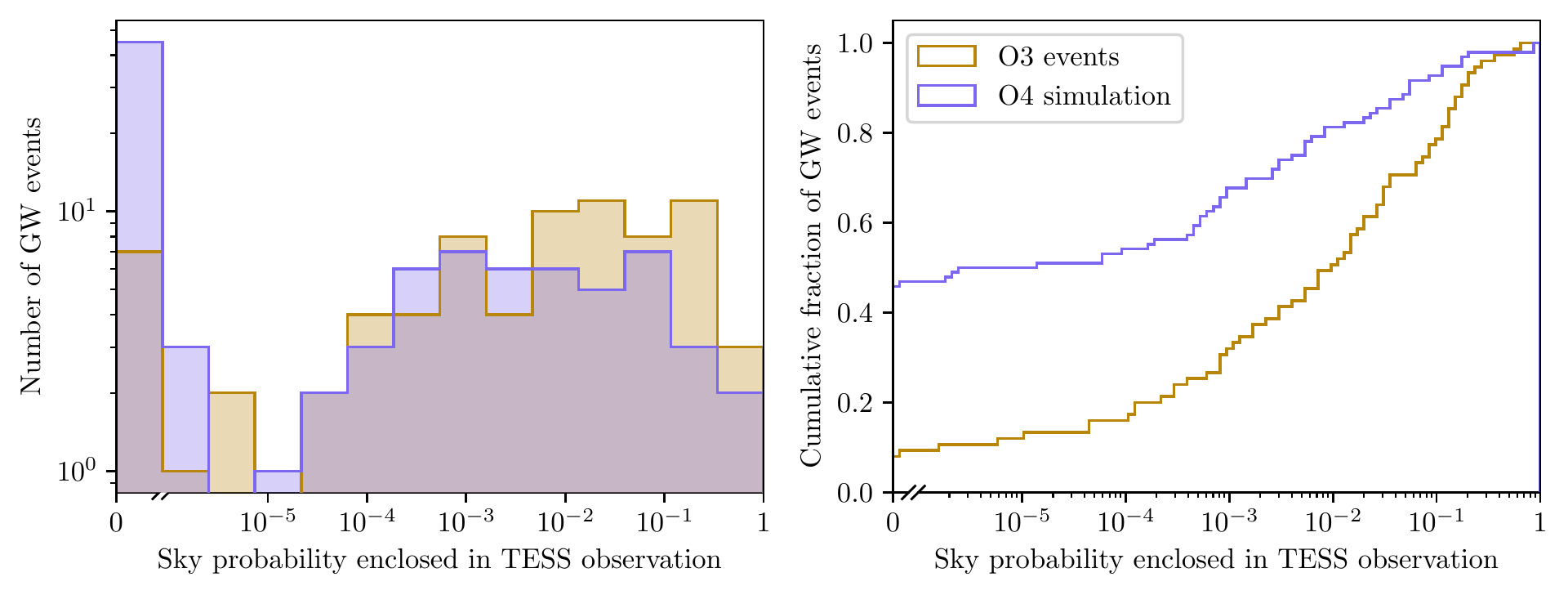}}
    \caption{Histogram \textit{(left)} and cumulative distribution \textit{(right)} of \ac{gw} skymap probability enclosed in \ac{tess} for O3 events and for the O4 \ac{bns} simulation.
    The O3 distribution includes all 75 events (\ac{bbh}, \ac{nsbh}, and \ac{bns}) publicly released by the LVK as part of GWTC-2 and -3.
    If the \ac{gw} event happened while \ac{tess} was not observing (7 of the 75 events), the overlap integral was calculated for the next available set of observations.
    The O4 distribution is for the 96 simulated \ac{bns} events.
    Because the simulated O4 events tend to have smaller \ac{gw} localization areas due to the improved \ac{gw} sensitivity and enlarged detector network, the chance of \ac{tess} observing part of the skymap is reduced compared to the O3 events.
    This is most evident in the cumulative distribution, where the fraction of O4 simulated events for which none of the skymap is covered by \ac{tess} is more than double that of the O3 events.
    At the other end of the distribution, where almost all of the skymap probability is covered by \ac{tess}, the gap between the O3 events and the O4 simulation narrows.
    This is because well-localized events have more concentrated areas of probability; either they are mostly covered by \ac{tess} or not at all.
    }
    \label{fig:enclosed_prob_dist}
\end{figure*}

For each of the 75 events from O3 in GWTC-2 and -3, we construct and calculate the overlap with the \ac{tess} \ac{fov} for the concurrent and subsequent sectors, up to $\sim 50$ days after the trigger (an example plot is shown in the left panel of Figure \ref{fig:example_skymap}).
Then, we selected \ac{gw} events where $>1$\% of the probability was enclosed within the \ac{tess} field of view of the concurrent or subsequent sector(s); this reduced the number of events we analyzed to 50. Of these, 33 had $>1$\% of the probability enclosed in the concurrent sector, while 17 had that amount of probability enclosed in a subsequent sector. 
Both cases were treated identically, in terms of the search technique.
The distribution of skymap probability enclosed in \ac{tess} for the 75 \ac{bns}, \ac{nsbh}, and \ac{bbh} events found in \acp{gw} in O3 is shown in Fig.~\ref{fig:enclosed_prob_dist}.

\subsubsection{Transient Identification}
\label{sec:trans-id}
In order to identify transient candidates that could correspond to an \ac{em} counterpart to the \ac{gw} trigger, we used a pipeline developed specifically to find transients in \ac{tess} \acp{ffi} (Jayaraman et al. in prep). 
This pipeline produces difference images using the ISIS package \citep{isis-citation} based on a median image constructed from $\sim$ 20 individual \acp{ffi} with low backgrounds. 
It calculates a root-mean-square (RMS) image from these, then finds strongly time-varying sources in this image using Source Extractor \citep{sextractor}. 
These sources are matched against the Gaia eDR3 catalog \citep{gaia-mission, gaia-edr3} and discarded if they are brighter than 19.5 in the G$_{\rm RP}$ band and have a parallax that is not consistent with zero (as that would likely correspond to a known stellar source).
We then filter the remaining sources through a convolutional neural network which has been trained on cutouts from the RMS images of \ac{tess} transients that were discovered by other observatories and surveys during prior sectors. 
This network assigns a probability for each source that captures its likelihood of being a bona fide transient. 
The pipeline constructs light curves for sources that have been assigned a probability greater than 0.6.
This results in a set of $\mathcal{O}(\gtrsim 1000)$ light curves per sector. 
The remaining light curves are still too many for a human to review in order to triage them for follow-up, so we make use of a light curve clustering algorithm that uses a random forest and the HDBSCAN algorithm \citep{hdbscan-campello} to identify anomalous light curves that could correspond to transients.
Clusters of light curves correspond to typical variability patterns, such as stellar variability (pulsations and eclipses) from nearby stars, scattered light patterns, and instrumental effects inherent to the CCD, such as hot pixels.
We review these final light curves by eye and study the most promising candidates in further detail.
An analysis of the non-kilonova transients detected by our pipeline and a more rigorous estimate of its limits will be discussed in a forthcoming paper.

\subsubsection{O3 search results}
\ac{gw} skymaps are composed of \texttt{HEALPix}\footnote{\url{http://healpix.sf.net/}} \citep{2005ApJ...622..759G, 2019JOSS....4.1298Z} pixels, which have limited resolution compared to \ac{tess}.
\ac{tess} pixels are 21'' on a side, while the highest resolution pixels in \ac{gw} skymaps average 52'' on a side, or $\sim2.5$ \ac{tess} pixels (this corresponds to a value of 4096 for the \texttt{HEALPix} NSIDE parameter), but most have significantly lower resolutions.
We use the \texttt{ang2pix} function in \texttt{healpy} to match each \ac{tess} transient to its corresponding \ac{gw} \texttt{HEALPix} skymap pixel.

We extracted $\lesssim 500$ possible light curve matches per \ac{gw} event for most of the considered events. 
For some events, we found over 1000 matching \ac{tess} transient triggers.
However, the vast majority of these are false positives, due to subtraction artifacts and nearby variable stars.
We discarded light curves corresponding to (a) instrumental noise (high RMS scatter or discontinuities in flux measurements), (b) long-period variability associated with any rotation or pulsation in a known nearby variable star (within $<$ 35''), or (c) consistency with the flux signature of a cataclysmic variable outburst (these targets were cross-referenced with existing CV catalogs, such as that in \citealt{2014MNRAS.441.1186D}). 
Additionally, we discarded light curves that matched faint stars in Gaia EDR3.
After the visual examination, we identified 76 (of the total $\sim$ 10 000) light curves associated with \ac{gw} event skymaps that were not filtered based on the above criteria. 
These 76 light curves had rises in flux starting $\sim$10--20 days after the associated \ac{gw} event, with rises spanning tens of days.

We investigated these rises in flux to ensure that they were not merely instrumental effects or contamination from nearby variable stars. 
To evaluate the possibility of instrumental systematics, we examined a 7\,px $\times$ 7\,px cutout from the raw \ac{ffi} around the source in question; single-pixel effects are strongly localized, and do not appear in nearby pixels in these images.
We also plotted all sources from Gaia EDR3 within 3 arcmin of each detected source; if there was a bright star (G$_{\rm RP} \lesssim$ 13.5) within 2 \ac{tess} pixels whose signal clearly bled into the central pixel of the cutout, that candidate source was discarded.
After filtering based on these criteria, we were left with 17 candidates that merited further investigation. 

We then searched existing catalogs using the NASA/IPAC Extragalactic Database\footnote{Accessible at \dataset[10.26132/NED1]{http://dx.doi.org/10.26132/NED1}.}---including the WISE point source catalog \citep{allwise-catalog}, GALEX point source catalog \citep{galex-catalog-paper}, and the SDSS galaxy catalog \citep{sdss-dr-17}.
We also queried the Pan-STARRS1 image cutout server and point source catalog from the MAST database \footnote{This data can be found in MAST: \dataset[10.17909/s0zg-jx37]{http://dx.doi.org/10.17909/s0zg-jx37}.} \citep{Chambers:2016jzn} to determine whether there was a galaxy at that location. 
Eight of the the 17 remaining light curves corresponded to WISE or GALEX point sources; these, along with the remaining nine sources, were ruled out as candidates due to the presence of a faint Gaia star (G$_{\rm RP}$ dimmer than 13.5) in each of the 40'' $\times$ 40'' Pan-STARRS cutouts around the coordinates of the light curves.
To further verify all the cases in which we suspected contamination from Gaia stars in the vicinity, we downloaded the available \ac{ffi} light curves for these stars from MAST and plotted them using the {\tt lightkurve} package \citep{lightkurve}.
These light curves were identical in morphology to the sources from the pipeline, demonstrating that the light curves found arise from contamination. 

Our pipeline was thus unable to identify any \ac{em} counterparts to \ac{bbh} mergers in the region of the \ac{gw} skymaps that overlaps the \ac{tess} \ac{fov}.
There is no detection of any obvious counterpart in the \ac{tess} \ac{fov} peaking above 17th magnitude in the \ac{tess} band. 
However, work is ongoing to more fully characterize the performance and detection limits of our transient pipeline, including deeper searches with lengthier stacks such as those presented in the light curve simulation section (Sec.~\ref{subsubsec:lcs}) below. 
Our work represents one of the first attempts to conduct a ``blind'' search for transients fainter than 17th magnitude in \ac{tess}.
Prior work in this arena \citep{tingay-frb-tess} centered on a specific goal---to identify optical counterparts to fast radio bursts in \ac{tess}'s \ac{fov}. 
\citeauthor{tingay-frb-tess} established a non-detection limit at 16th magnitude.
However, by broadening this search to all classes of transients (including counterparts to compact object mergers), 
\ac{tess} can cast a wider net and possibly answer lingering questions about transient physics across a wide range of transient events.

We paid particular attention to the events which included at least one neutron star and thus are more likely to result in \ac{em} emission: the \ac{bns} GW190425, the confident \acp{nsbh} GW191219\_163120, GW200105\_162426, and GW200115\_042309, and the potential \acp{nsbh} GW190814 and GW200210\_092254.
Counterparts from these events such as kilonovae are expected to peak within 1--2 days of the \ac{gw} trigger, so we focused on the \ac{tess} sectors that were concurrent with the \ac{gw} merger time.
Unfortunately, \ac{tess} did not overlap significantly with the skymaps for these events, with the most considerable overlap being 3.9\% of the sky probability for the \ac{nsbh} GW191219\_163120 during sector 19.
The \ac{nsbh} events represented 12 of the 76 light curves that were flagged during our visual inspection.
Further analysis of these light curves did not reveal anything consistent with kilonova light curves. 
One event, GW200210\_092254, did have a light curve associated with its localization that had a rise and fall during the \ac{tess} observation. 
However, the start of emission occurs $\sim$ 15 days after the \ac{gw} trigger, much longer than what most kilonova models predict.
Additionally, the coordinates of the \ac{em} transient lie at the 99.8\% credible level of the \ac{gw} localization area, indicating that it is highly unlikely to be associated with GW200210\_092254\footnote{
The smallest credible levels of a skymap include only the most likely pixels, while larger credible levels also include less likely pixels.
}.
Thus, we claim no kilonova detections from the overlapping region of \ac{tess}'s \ac{fov} and the \ac{gw} skymaps from O3, down to at least a limiting magnitude of $\sim$ 17.
Future work may strengthen these constraints.

\subsubsection{SDSS J1249+3449}

In addition to the ``blind'' search for \ac{em} transients from O3 (discussed above), we also studied the \ac{tess} light curve of the quasar SDSS J124942.30+344928.9.
\citeauthor{2020PhRvL.124y1102G} argued that a flare observed in this galaxy (referred to as J1249) in 2019 June might correspond to a \ac{bbh} merger product being kicked upward through the accretion disk of an \ac{agn}.
TESS observed J1249 in Sectors 22 (2020 Feb--March) and 49 (2022 Feb--March).

\begin{figure}
\centering
    \includegraphics[width=\linewidth]{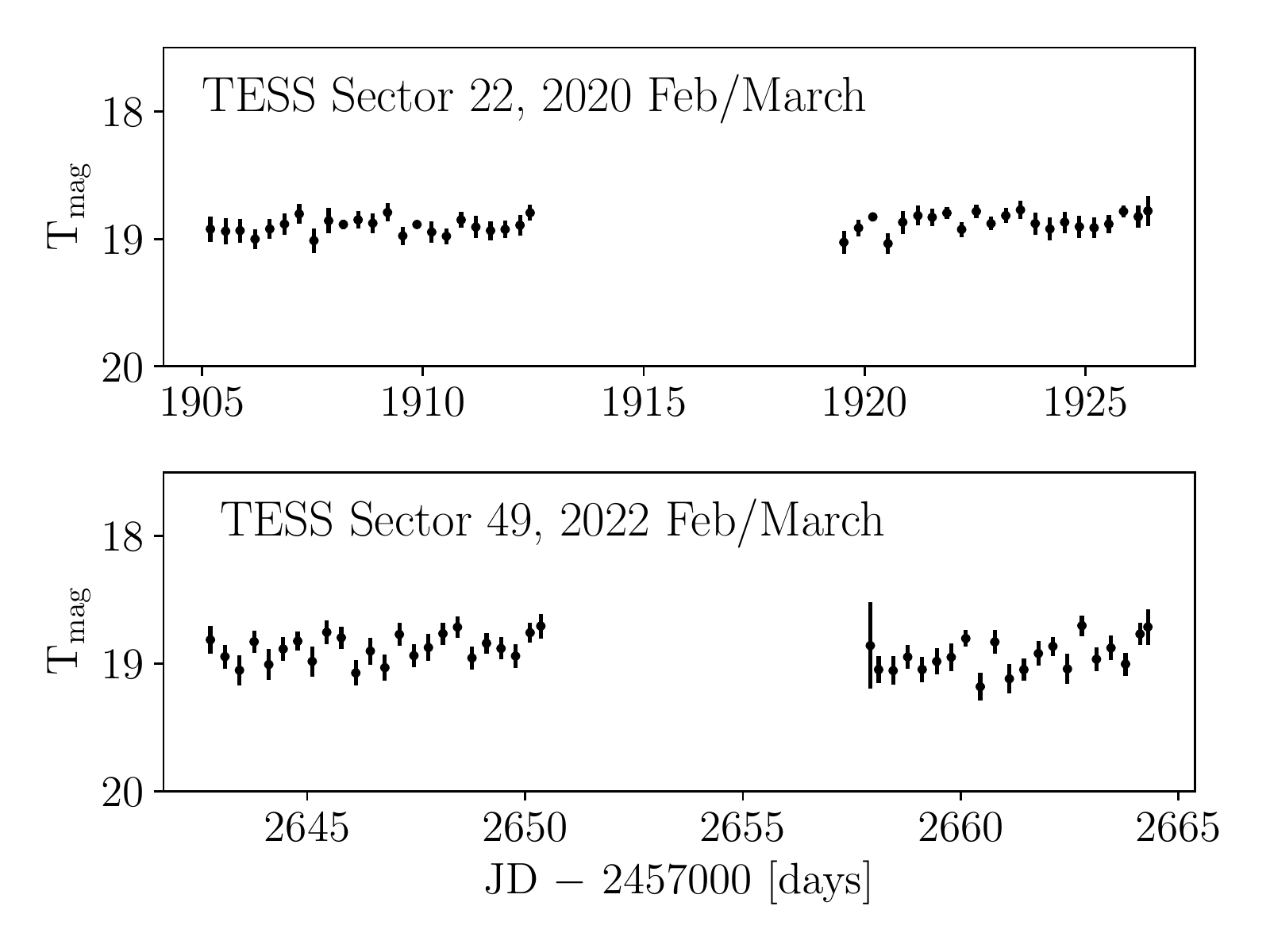}
    \caption{\ac{tess} light curves of SDSS J1249+3449, the \ac{agn} with a flare reported by \citeauthor{2020PhRvL.124y1102G}. 
    The light curves are binned to 8 hours, and the error bars show the standard deviation in each bin scaled by the square-root of the number of data points. 
    In Sector~22, the light curve is consistent with a constant light curve. In Sector~49, there is excess variability at the 5\% level, but inspection of the difference images suggests that these are residual systematic errors.
    }
    \label{fig:sdss1249}
\end{figure}

We extracted light curves at the position of J1249 from the \ac{tess} \acp{ffi} using forced photometry, as described in \citet{Fausnaugh2021}. 
The results are shown in Figure~\ref{fig:sdss1249}.
After correcting for measurement uncertainties \citep{Nandra1997,Vaughan2003}, the light curve from Sector~22 is consistent with a constant light curve on timescales longer than 8 hours. 
For Sector~49, the light curve is flat but exhibits a residual scatter of 5.8\% on 8 hour timescales after correcting for measurement uncertainties. On shorter timescales than 8 hours, there is formally more observed variability than can be explained with statistical uncertainties at the 2\% level. 
However, inspection of the difference images shows that the variability is not associated with the \ac{agn} and so is likely due to residual systematic errors or a nearby variable star. 
These systematic errors likely contribute to the excess variance at 8 hour timescales.

The Sector~49 observations were approximately three years after the original flare in 2019 and so are consistent with the expected timescale for the re-entry of the \ac{bbh} merger remnant into the \ac{agn} disk. 
However, the short \ac{tess} baseline of 27~days is poorly suited to identify flares lasting 50~days or longer, especially considering uncertainties in the delay between the merger event and ejection of the remnant from the disk \citep{Graham2022}, as well as uncertainty in the orbital period of the remnant. 
We therefore cannot put a meaningful constraint the \ac{bbh} merger hypothesis for J1249 with these \ac{tess} observations.

\section{Simulated performance in O4} 
\label{sec:o4sim}
During the \ac{lvk}'s next observing run, O4, the increased sensitivity of \ac{gw} detectors and the addition of KAGRA to the overall detector network will mean that \ac{gw} events are expected multiple times per week, and potentially even daily \citep{Aasi:2013wya}.
The majority of these events will be \ac{bbh} mergers, with a comparatively smaller number of \ac{bns} and \ac{nsbh} mergers \citep{Abbott:2020qfu}.
In this section, we evaluate the prospects of detecting a kilonova in \ac{tess}.

\subsection{Simulated BNS population and light curves} \label{subsec:bnses}
To simulate a population of kilonovae in O4, we use the \ac{bns} simulations released in \citet{Frostig:2021vkt}, which include 625 neutron star mergers with realistic mass and spin distributions. 
In these simulations, \ac{bns} mergers are placed uniformly in comoving volume out to a luminosity distance ($d_L$) of 400 Mpc and isotropically positioned on the sky, and are distributed randomly in time over the 2022 calendar year.\footnote{
At the time that these simulations were created, O4 was slated to begin in early 2022. 
Our study is intended to examine \ac{tess}'s utility for identifying \ac{gw} counterparts in general, not just for a specific set of \ac{tess} pointings in a given time period. 
Thus, using 2022 \ac{tess} pointings for events simulated to be during 2022 should not significantly affect our results.
}

Each event is then associated with a GW detector configuration, assuming a duty cycle of 70\% (i.e., each detector has a 70\% chance of being in observing mode) and added into Gaussian noise.
The \pycbc matched-filter GW search algorithm is then used to recover each BNS using detector power spectral densities (PSDs) representative of O4 sensitivities \citep{Nitz:2018rgo, DalCanton:2020vpm, noise_curves}.
We consider an event to be detected in \ac{gw}s if \pycbc recovers it with a network \ac{snr} greater than 9. 
Each detected event is also associated with a localization from the full \bilby parameter estimation pipeline \citep{Ashton:2018jfp, Romero-Shaw:2020owr}; see \cite{Frostig:2021vkt} for further details on the simulation.

In our analysis, we choose astrophysical \ac{bns} merger rates of 50, 250, and 1000 Gpc$^{-3}$ yr$^{-1}$; these represent a broad range that spans the current uncertainties from various population models presented in \citet{LIGOScientific:2021psn}.
For each rate, we calculate the total number of mergers with $d_L < 400$\,Mpc in one year (e.g., 52 events for the 250 Gpc$^{-3}$ yr$^{-1}$ rate).
We then randomly draw with replacement that number of events from the set of 625.
Each event is associated with a light curve and a determination of whether or not it is detected in \acp{gw}.
This process is repeated 100 times, each with a different random seed, for each rate to build a distribution of possible results for O4.

For each of the 625 events, we generate simulated kilonovae light curves using the \citet{Kasen:2017sxr} models for the \ac{tess} bandpass.
These models are computed by solving for relativistic radiation transport in radioactive plasma, with the assumption of a spherical kilonova geometry. 
The models have three parameters: the mass of the dynamical ejecta $M_{\rm{ej}}$, the expansion velocity of that ejecta $v_{\rm{ej}}$, and the lanthanide fraction of the ejecta $X_{\rm{lan}}$.
We use fitting formulae from \citet{Coughlin:2018fis} to determine $M_{\rm{ej}}$ and $v_{\rm{ej}}$ from the primary and secondary masses ($m_1$ and $m_2$) simulated for each event. 
The lanthanide fraction $X_{\rm{lan}}$ remains a free parameter; since $X_{\rm{lan}}$ for binary neutron star mergers remains poorly understood, we choose four values of $X_{\rm{lan}}$ to generate light curves: $10^{-2}$, $10^{-3}$, $10^{-4}$, and $10^{-5}$.

We then run the overlap tool described in Sec.~\ref{subsec:overlap} for the \ac{gw} \bilby skymaps corresponding to each of the events found in gravitational waves.
Figure~\ref{fig:enclosed_prob_dist} shows the distribution of \ac{tess}-enclosed probabilities for the simulated O4 \ac{bns} events.
The distribution is roughly similar to the distribution of detected \ac{bbh}, \ac{nsbh}, and \ac{bns} events from O3, with slightly less probability enclosed for the simulated O4 events due to the improved localizations compared to O3 (and therefore more reliant on serendipitous observation for any significant coverage).

\subsection{Simulation results} \label{subsec:results}

\begin{figure}[ht!]
\centering
	\includegraphics[width=\linewidth]{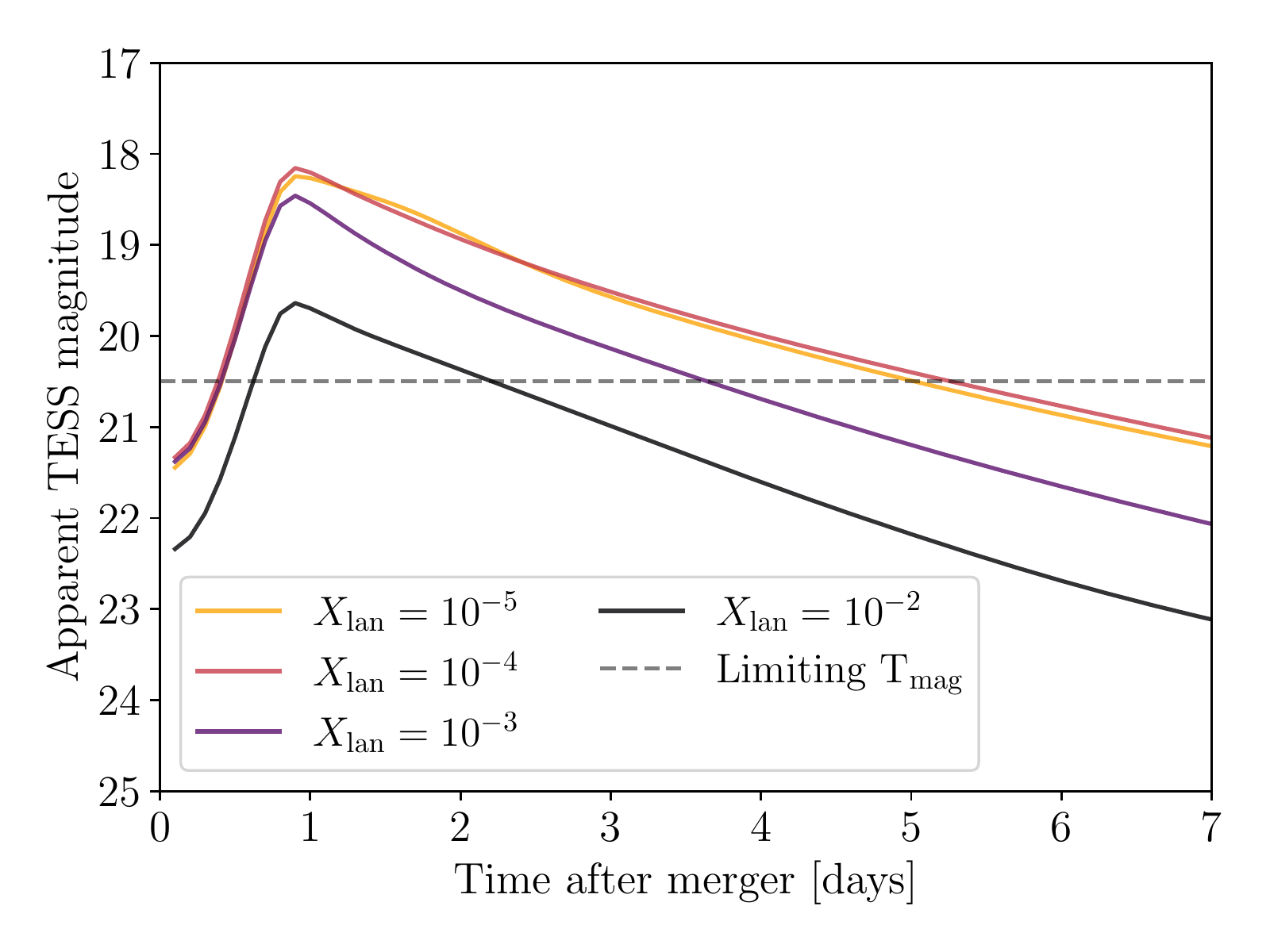}
    \caption{Example \ac{tess}-band kilonova light curves for one of the brightest \ac{gw}-detected events in the simulation, for a \ac{bns} merger at a luminosity distance of 64 Mpc.
    Light curves for various values of the lanthanide fraction ($X_{\rm lan}$) are shown, along with an estimate for the 3-$\sigma$ \ac{tess} limiting magnitude of a 4-hour stack ($T_{\rm mag} \sim 20.5$).
    Ejecta with higher lanthanide fractions have lower opacity, which obscures the optical and near-infrared emission.
    Since the \ac{tess} bandpass is in the optical and near-infrared, it is more sensitive to kilonovae with lower lanthanide fractions.
    For all lanthanide fractions, \ac{tess} should be able to observe at least part of the rise and decay of the light curve of a sufficiently bright kilonova.
    }
\label{fig:example_lc_xlans}
\end{figure}

\subsubsection{Light curves} \label{subsubsec:lcs}

In Figure~\ref{fig:example_lc_xlans}, we show results for the simulated light curves in the \ac{tess} band across various lanthanide fractions, for one of the brightest \ac{gw}-detected events in the simulation.
We find that the lowest lanthanide fractions $X_{\rm{lan}}$ of $10^{-4}$ and $10^{-5}$ lead to very similar light curves, which are the brightest in the \ac{tess} band.
The $X_{\rm{lan}} = 10^{-3}$ curve is slightly less luminous by $\sim 0.25$\,mag at its peak; this difference increases to nearly one full magnitude in the tail of the light curve.
The $X_{\rm{lan}} = 10^{-2}$ light curve is significantly dimmer: it peaks about two magnitudes lower than the $X_{\rm{lan}} = 10^{-5}$ and $X_{\rm lan} = 10^{-4}$ models.
These results are caused by the large number of line transitions for lanthanides which lead to high opacities \citep{Kasen:2013xka}.
This opacity acts to obscure optical emission at higher lanthanide fractions, resulting in relatively stronger infrared emission.
While \ac{tess}'s passband does receive some infrared signal (it is broad-band from 600 to 1000 nm; see Fig.~\ref{fig:bands}), the bulk of its passband is in the optical and near-infrared.
This makes it comparatively better suited for observations of kilonovae with lower lanthanide fractions.
GW170817, which \citet{waxman-xlan} suggest had $X_{\rm lan} \sim 10^{-3}$ (corresponding to the purple light curve in Figure \ref{fig:example_lc_xlans}), would have been detected by \ac{tess} had it been observing that region of sky.

\begin{figure}[ht!]
\centering
	\includegraphics[width=\linewidth]{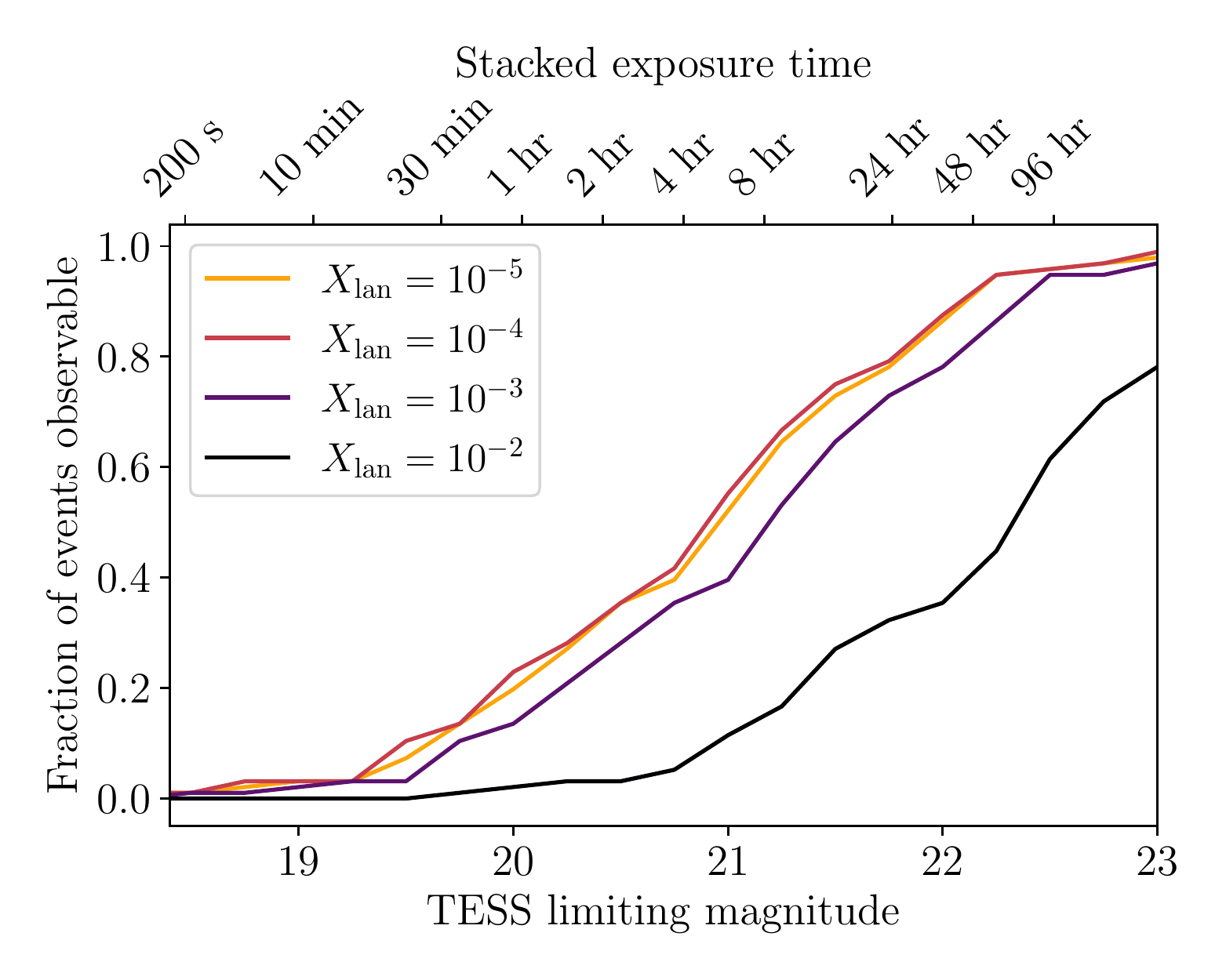}
    \caption{Distribution of the fraction of \ac{gw}-detected events observable as a function of the overall stacked integration time and (top axis) and \ac{tess} limiting magnitude (bottom axis), at a variety of lanthanide fractions.
    Stacking \ac{tess} \acp{ffi} lowers the limiting magnitude, allowing us to detect more events, especially for kilonovae with lower lanthanide fractions.
    For higher lanthanide fractions (e.g., $X_{\rm{lan}} = 10^{-2}$), over three-quarters of events will not be detectable at an 8 hour stack ($\sim$ 21 mag); the majority would require an \ac{ffi} stack of a day or more to be detectable.
    Lower lanthanide fractions, on the other hand, will result in almost 60\% of \ac{gw}-detected events being bright enough to be detectable using an $\sim$8 hour stack of \ac{ffi}s.
    Note that the limiting magnitude as a function of stacked exposure time presented here is based only on counting noise on the source and background.
    In practice, the limiting magnitude can vary significantly, by up to 2-3 magnitudes, due to image backgrounds, proximity of targets to other bright objects, and other factors. 
    }
\label{fig:lim_mag_observable}
\end{figure}

We note that \ac{tess}'s limiting magnitude, and thus its sensitivity to kilonovae, is closely related to the timescale at which the light curve is binned.
During \ac{tess}'s \ac{em2}, the stacked integration time of each \ac{ffi} will be 200 seconds; consequently, stacking the \acp{ffi} (or, similarly, binning the resulting light curves) will likely be necessary to reveal faint transients.
Figure~\ref{fig:lim_mag_observable} shows the fraction of events observable as a function of \ac{tess} limiting magnitude and stacked exposure time.
The mapping between stacked exposure time and limiting magnitude was done based on (a) the counting uncertainty in background-dominated \acp{ffi} being 0.409 e$^{-}$/s (assuming an aperture of 4 pixels) and (b) the zeropoint of \ac{tess} being 20.44 (i.e., this is the magnitude corresponding to a count rate of 1 e$^{-}$/s)\footnote{
These figures are from the \ac{tess} Instrument Handbook at \url{https://archive.stsci.edu/files/live/sites/mast/files/home/missions-and-data/active-missions/tess/_documents/TESS_Instrument_Handbook_v0.1.pdf}
}.

If \ac{bns} ejecta have lanthanide fractions as high as $X_{\rm{lan}} = 10^{-2}$, stacking \acp{ffi} to 8 hours (reaching a limit of $\sim$21 mag) will reveal less than 20\% of \ac{bns} mergers found in \ac{gw}s.
At lower lanthanide fractions, almost half of \ac{bns} mergers in the \ac{tess} \ac{fov} will be bright enough to be detectable with an 8-hour stack; with a 24-hour stack, nearly all such mergers are detectable. 

However, when we stack images or bin light curves, we experience a tradeoff between limiting magnitude and the number of useful data points.
For instance, stacking \acp{ffi} over many days can push limiting magnitudes down to 21.5 (or even slightly fainter---\citealt{rice-stack-limits}, for example, reach 21.8), but will only provide one data point every few days with which to identify kilonovae.
Since kilonovae rise and fade on the order of days, this can make discerning kilonovae at fainter magnitudes more difficult, if not impossible.
These binned light curves could also be confused with other, non-kilonova signals; consequently, it is crucial to carefully choose the stacking or binning timescale to reach fainter magnitudes while not washing out any useful signal.
A minimum stacked integration time of 30\,min (i.e., stacking 9 200-s \ac{ffi}s to a limiting magnitude of 19.6) is necessary to detect any substantial fraction of kilonovae.
For shorter stacks, only a few percent of \ac{gw}-found \ac{bns} mergers result in \ac{em} emission that will be detectable in \ac{tess}.
These estimates are also all under the optimistic expectation that the backgrounds in the light curves are low; in certain sectors, scattered light can cause high backgrounds in the \acp{ffi}.
In those cases these detection limits would become brighter, closer to 19th magnitude for an 8-hour stack (see, e.g., \citealt{faus-ia-sne}).

\begin{figure}[ht!]
    \centering
	\includegraphics[width=\linewidth, ]{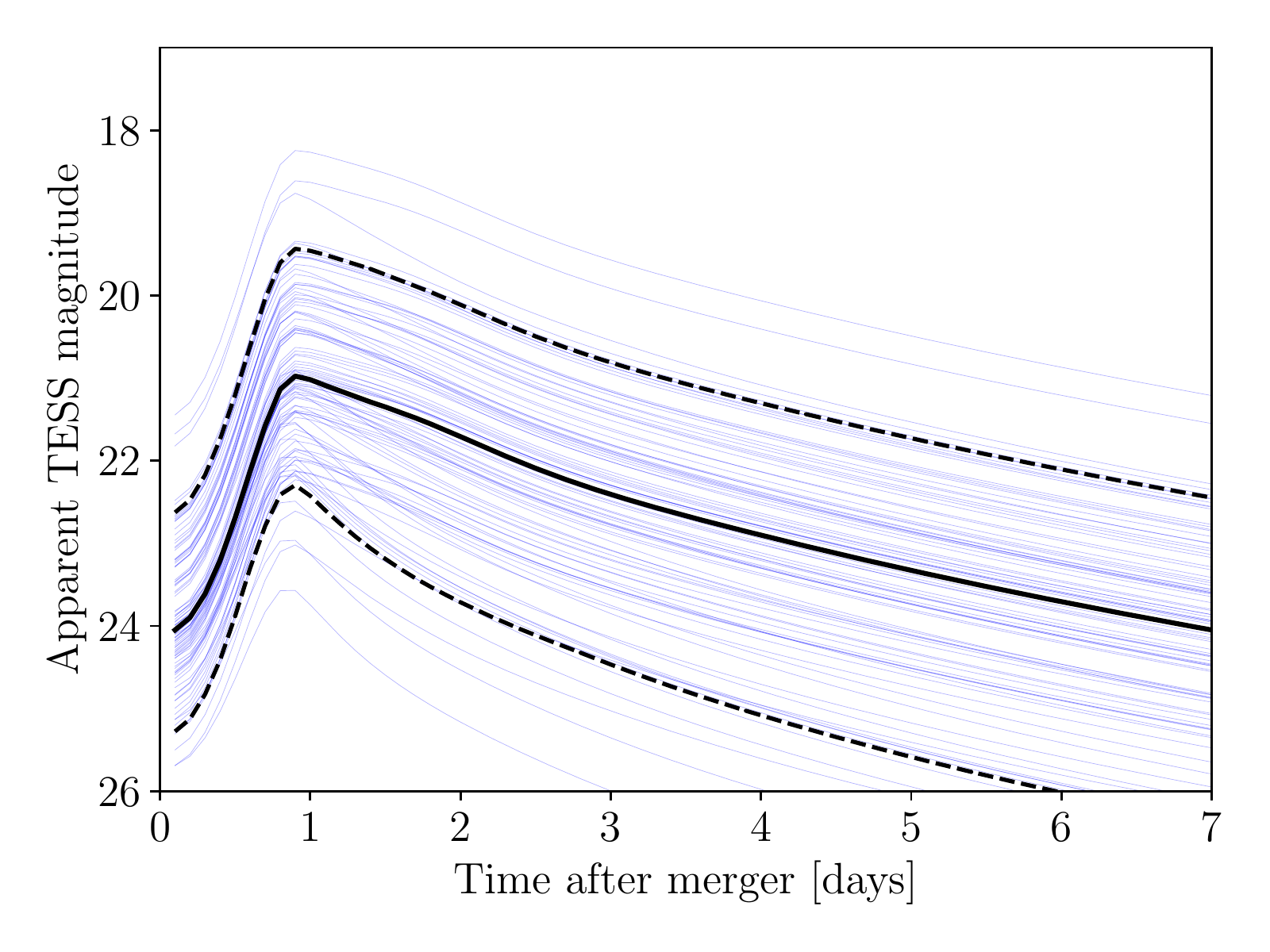}
    \caption{\ac{tess}-band light curves for all 96 simulated O4 \ac{bns} events found in gravitational waves for $X_{\rm{lan}} = 10^{-5}$.
    Individual light curves are shown in light blue. 
    The median light curve is in solid black, and the 90\% interval is bounded by the black dotted lines.
    The brightest light curves can peak at almost 18 mag, whereas the median light curve peaks near 21 mag.
    Despite varying dynamical ejecta mass $M_{\rm{ej}}$ and ejecta expansion velocity $v_{\rm{ej}}$ across the simulated events, the light curves tend to all peak at $\sim 1$ day after merger and subsequently decay at a similar rate.
    }
    \label{fig:all_lcs}
\end{figure}

Figure~\ref{fig:all_lcs} shows light curves in the \ac{tess} band for all 96 \ac{bns} events out of the 625 simulated events which were detected in \ac{gw}s, assuming an $X_{\rm{lan}} = 10^{-5}$.
The median light curve peaks at just fainter than 21 mag, with the brightest 5\% peaking between 18 and 19 mag. 
The light curves tend to fade at a similar rate of 0.5 mag/day regardless of the values of the ejecta mass $M_{\rm ej}$ and velocity $v_{\rm ej}$.
This suggests that the vast majority of kilonova detections in \ac{tess} will be aided by stacking \ac{ffi}s and/or binning light curves to probe these sources that are at the limit of \ac{tess}'s sensitivity.

\subsubsection{Events in TESS} 
\label{subsubsec:intess}
Next, events detected by \pycbc in gravitational waves are checked to see if they are within the \ac{tess} \ac{fov}.
The number of \acp{bns} identified in gravitational waves over a year of observations during O4 ranges from a mean of 1.5 for the most pessimistic rate to 32 at the most optimistic (90\% confidence interval), where we have averaged over our 100 random seeds.
While we cannot establish a single limit for detectability in \ac{tess} due to the variable backgrounds in different sectors, we can evaluate the number of kilonovae detected at various limiting magnitudes. 
These limiting magnitudes, as discussed above, correspond to different stacking times for the \acp{ffi}.
Table~\ref{tab:found} gives the number of events detected in \ac{gw}s,  covered by the \ac{tess} \ac{fov}, and bright enough to be detectable by \ac{tess} for each merger rate at various limiting magnitudes.

\begin{table*}
\centering
\caption{Results for BNS mergers found in gravitational waves.
We present the astrophysical BNS rate, the number of events found in gravitational waves by a matched-filtering search, the number of events which are in the \ac{tess} \ac{fov}, and of those, the number of events which have simulated \ac{tess} light curves brighter than a limiting magnitude of 21.
These light curves assume a $X_{\rm{lan}} = 10^{-5}$.
In this simulation, using reasonable stacked integration times to probe deeper than a limiting magnitude of 21 does not result in any change in the number of observable kilonovae.
For the ``Found,'' ``Covered,'' and ``Bright'' columns, we report the mean over 100 random seeds, as well as the median, 5th, and 95th percentiles in parentheses.
}
\label{tab:found}
\begin{tabular}{cccc}
\hline
\multicolumn{1}{c}{BNS rate} & 
\multicolumn{1}{c}{Found} & 
\multicolumn{1}{c}{Covered} &
\multicolumn{1}{c}{Bright at} \\
\multicolumn{1}{c}{(Gpc$^{-3}$yr$^{-1}$)} &
\multicolumn{1}{c}{in GWs} &
\multicolumn{1}{c}{by TESS} &
\multicolumn{1}{c}{limiting mag 21} \\
\hline
50 & 1.5 ($1^{+2}_{-1}$) & 0 ($0^{+0}_{-0}$) & 0 ($0^{+0}_{-0}$) \\
250 & 8.5 ($8^{+5}_{-3}$) & 0.2 ($0^{+1}_{-0}$) & 0.1 ($0^{+1}_{-0}$) \\
1000 & 33 ($33^{+9}_{-7}$) & 0.7 ($0^{+3}_{-0}$) & 0.4 ($0^{+2}_{-0}$) \\
\hline
\hline
\end{tabular}

\end{table*}
 
These results for the number of events covered in \ac{tess} are in line with a simple order of magnitude calculation: 
both the detected \ac{bns} events and the \ac{tess} coverage during \ac{em2} are roughly isotropic across the sky, and \ac{tess} at any given observation covers $\sim$5.6$\%$ of the sky.
Taking the median value of the most optimistic rate (32 events in \acp{gw}), multiplied by this percentage, results in 1.8 events covered in the \ac{tess} \ac{fov}; this value is within the uncertainty of the value found in the simulation (mean of 0.4, with median and 5th, 95th percentiles of $0^{+3}_{-0}$).

Figure~\ref{fig:enclosed_prob_dist} shows the distribution of probability enclosed by \ac{tess} for the 96 \ac{bns} events simulated and found in \ac{gw}s.
The simulated performance at enclosed probabilities larger than 10\% is similar to that of the 75 \ac{bns}, \ac{nsbh}, and \ac{bbh} mergers observed in O3, with approximately 25\% of events having greater than 5\% of their probability enclosed in \ac{tess}.

\subsection{Subthreshold simulation results} \label{subsec:subthresh_results}
\begin{table*}
\centering
 \caption{Results for subthreshold BNS mergers \textit{not} found in gravitational waves.
Same as in Table~\ref{tab:found}, but for subthreshold events which are below the \ac{gw} search algorithm's signal-to-noise ratio threshold and thus deemed to not be found in gravitational waves.
Here, we also show results for various limiting magnitudes which can be obtained with stacked integration times.
}
\label{tab:subthresh}
\begin{tabular}{cccccc}
\hline
\multicolumn{1}{c}{BNS rate} & 
\multicolumn{1}{c}{Found} & 
\multicolumn{1}{c}{Covered} &
\multicolumn{3}{c}{Bright at limiting mag}  \\
\cline{4-6}
\multicolumn{1}{c}{(Gpc$^{-3}$yr$^{-1}$)} &
\multicolumn{1}{c}{in GWs} &
\multicolumn{1}{c}{by TESS} &
\multicolumn{1}{c}{21.5} &
\multicolumn{1}{c}{21} &
\multicolumn{1}{c}{20.5} \\
\hline
50 & 0 & 0.6 ($0^{+2}_{-0}$) & 0.2 ($0^{+1}_{-0}$) & 0.2 ($0^{+1}_{-0}$) & 0.1 ($0^{+1}_{-0}$) \\
250 & 0 & 2.7 ($3^{+2}_{-3}$) & 1.1 ($1^{+2}_{-1}$) & 0.7 ($1^{+1}_{-1}$) & 0.6 ($0^{+2}_{-0}$) \\
1000 & 0 & 11.0 ($11^{+5}_{-4}$) & 4.6 ($5^{+3}_{-4}$) & 3.0 ($3^{+3}_{-3}$) & 2.2 ($2^{+3}_{-2}$) \\
\hline
\hline
\end{tabular}

\end{table*}

Thus far, we have discussed searching \ac{tess} data with a \ac{gw}-triggered search.
\ac{tess}'s wide \ac{fov} and continuous observation capabilities mean that it can also be used to conduct a search for \ac{bns}s in the opposite direction, with potential kilonovae in \ac{tess} triggering subsequent searches in \ac{gw} data, similar to searches performed in ZTF \citep{Andreoni:2021ykx}.
Unlike ground-based surveys, \ac{tess} enjoys uninterrupted coverage that spans for weeks, with no periods of night; its pointing stability ($\lesssim 1\%$ of a pixel for a 1-hour stack)\footnote{These are shown in the sector-by-sector Data Release Notes.} is also such that stacking \ac{tess} \acp{ffi} is more straightforward than stacking ground-based images.
These properties mean that \ac{tess} is particularly well suited to searching for fast and faint transients such as kilonovae that may require significant stacking to achieve a confident detection.
As a proof of concept, we repeat the analysis in Sec.~\ref{subsubsec:intess} for ``subthreshold'' events, which were below the \pycbc \ac{snr} threshold and thus not detected in \ac{gw}s.
These results are presented in Table~\ref{tab:subthresh}. 

Even for the lowest \ac{bns} rate, we find that \ac{tess} could potentially identify one kilonova independent of \ac{gw} observations.
For the highest rate, we expect to observe means of between two and five \ac{bns} events, depending on the stacked integration time.
This emphasizes the importance of \ac{em}-triggered searches in \ac{gw} data, which trawl deeper into \ac{gw} detector noise.
Additional compact object mergers are statistically guaranteed to exist in the \ac{gw} data at lower \ac{snr}, and joint searches between the \ac{lvk} detector network and \ac{tess} will contribute towards recovering them.

\section{Discussion} \label{sec:discussion}
The results presented in Sec.~\ref{sec:o4sim} assume a \ac{tess} detection if the kilonova light curve peaks above the limiting magnitude for a given stack of \acp{ffi} (e.g., for a 4-hour stack, the limiting magnitude is 20.5).
This simplification is reasonable in the mode where a search is being conducted in \ac{tess}, based on a pre-existing \ac{gw} trigger with a skymap and a time. 
On the other hand, searches in \ac{tess} for subthreshold \ac{gw} events will necessarily be shallower, and multiple data points above the limiting magnitude will be needed to definitively classify a light curve as a signature of the \ac{em} emission arising from a compact object merger.
This assumption means that our subthreshold search results presented here are likely optimistic.
However, they still represent an important new synergistic use of \ac{tess} with \ac{gw} detectors in the \ac{lvk} network, and a kilonova discovery using \ac{tess}---independent of \ac{lvk}---would be significant.

In our analysis, we have also focused largely on the results for $X_{\rm{lan}} = 10^{-5}$, which typically lead to the brightest light curves in the mostly-optical \ac{tess} band.
Lower values of $X_{\rm{lan}}$ will likely lead to worse kilonova performance from \ac{tess}. 
\citet{waxman-xlan} provides an estimate of $X_{\rm lan} \sim 10^{-3}$ for GW170817.
However, they also note that this is a poorly constrained value, as there remain significant uncertainties in the values of the infrared opacities of heavy elements such as the lanthanides and their relative contributions to the opacity of kilonova ejecta (see, e.g., \citealt{Kasen:2013xka}).

Despite these caveats and the small number of expected detections in O4, \ac{tess} remains a powerful tool for multi-messenger astronomy, especially since these observations have no opportunity cost compared to other targeted follow-up programs.
\ac{tess}'s observing schedule is set well in advance, and unlike targeted follow-up, there are no targets that must be prioritized over others.
Furthermore, the potential treasure trove of \ac{tess} \acp{ffi} are publicly released on a weekly basis without any proprietary period, which makes them particularly useful for transient searches and follow-up.

If another GW170817-like event occurs in O4 and happens to be in \ac{tess}'s \ac{fov}, the kilonova should be bright enough that \ac{tess} will catch the rise of the light curve.
GW170817 peaked at about 17.5 mag in the i- and z-bands \citep{Drout:2017ijr}, which have overlap with the \ac{tess} bandpass (see Fig. \ref{fig:bands}).
Crucially, we will obtain a light curve from \ac{tess} with a data point every 200\,s.
The kilonova signal for such an event may be visible in the raw light curve already, but stacking the \ac{ffi}s to probe fainter magnitudes and using these stacks to construct a binned light curve may help us identify further trends in the light curve's rise.
\ac{tess} will join the ranks of new surveys such as BlackGEM \citep{Bloemen:2016vqg}, the Argus Array \citep{Law:2021jnn}, and others \citep{Chase:2021ood} in performing high-cadence, wide-field imaging of potential \ac{gw} counterparts.
These surveys can help constrain kilonova physics.

Finally, we can take advantage of \ac{tess}'s wide \ac{fov} to exclude large areas of the skymap to preserve telescope time for other observers. 
Since \ac{tess}'s observing and downlink schedules are known well in advance and processing times are consistent, we will be able to say at the moment of a \ac{gw} event whether or not \ac{tess} was concurrently observing any part of the probability skymap.
If so, we can expedite the processing of the data after downlink and focus our search on the highest-likelihood regions of the overlap using our transient pipeline.
This, along with possible integration into the TreasureMap online service \citep{2020ApJ...894..127W}, can help other observatories with smaller \acp{fov} plan their observing schedules to avoid redundancy with \ac{tess}. 
This will likely increase the chance of detecting a kilonova (or other \ac{em} counterpart) before it fades away forever.

\section{Conclusion} \label{sec:conclusion}
We have conducted an analysis of \ac{tess}'s capabilities for observations of \ac{em} counterparts to \ac{gw} events, with a study of archival \ac{tess} data from O3 and a simulation for \ac{tess}'s performance and potential contributions for kilonovae observations during O4, slated to begin in 2023 March.
We find no evidence of \ac{em} counterparts in \ac{tess} during O3 down to a magnitude of $\sim 17$ after inspection of all skymaps with more than 1\% sky probability that overlapped with the \ac{tess} \ac{fov}.
We also find no evidence for a re-entry flare from the \ac{agn} that \citeauthor{2020PhRvL.124y1102G} report as the site of an potential \ac{em} counterpart to the \ac{bbh} merger GW190521, during the two sectors that \ac{tess} observed it.

In O4, with the most optimistic \ac{bns} rate of 1000 Gpc$^{-3}$yr$^{-1}$ and a favorable $X_{\rm{lan}}$ of $10^{-5}$, we expect at most one detection in \ac{tess} of a \ac{bns} merger found in \ac{gw}s; however, lower rates and/or larger lanthanide fractions may lower this detection rate and make a concrete kilonova identification more unlikely.
On the other hand, we find that \ac{gw}-independent searches in \ac{tess} for kilonovae may be able to uncover up to five ``subthreshold'' \ac{bns} events which are buried in \ac{gw} detector noise (for the most optimistic \ac{bns} rate), highlighting the utility of \ac{em}-triggered searches in \ac{gw} data.

Finally, we discuss other aspects of \ac{tess} and its capabilities which make it a powerful tool for multimessenger astronomy.
Chief among these is that searches in \ac{tess} data do not require any additional planning or telescope time, as its observations of specific regions of the sky are made on a fixed schedule set years in advance.
Consequently, there is no tradeoff necessary between multi-messenger science and other observational interests, unlike other targeted follow-up programs for \ac{em} counterparts to \ac{gw} events.
Additionally, its wide \ac{fov} and continuous month-long observations make it one of the most ideal observatories for serendipitous observations of transients which might be missed by traditional ground-based surveys; besides the \ac{gw} events mentioned here, neutrinos and gamma ray bursts may also prove to be complementary targets.
We expect \ac{tess}'s EM2 and \ac{lvk}'s O4 to produce a treasure trove of data and observations that could significantly enhance our understanding of compact object mergers.

\acknowledgments{
We thank Sylvia Biscoveanu, Danielle Frostig, Viraj Karambelkar, Andr\'as P\'al, Leo Singer, and Michael Coughlin for helpful discussions and comments.

G.~M. and E.~K. acknowledge support of the National Science Foundation and the LIGO Laboratory.
LIGO was constructed by the California Institute of Technology and
Massachusetts Institute of Technology with funding from the National
Science Foundation and operates under cooperative agreement PHY-0757058.
This material is based upon work supported by NSF's LIGO Laboratory which is a major facility fully funded by the National Science Foundation.

This paper includes data collected by the TESS mission. Funding for the TESS mission is provided by the NASA Science Mission Directorate.

This research has made use of data or software obtained from the Gravitational Wave Open Science Center (gwosc.org), a service of LIGO Laboratory, the LIGO Scientific Collaboration, the Virgo Collaboration, and KAGRA. LIGO Laboratory and Advanced LIGO are funded by the United States National Science Foundation (NSF) as well as the Science and Technology Facilities Council (STFC) of the United Kingdom, the Max-Planck-Society (MPS), and the State of Niedersachsen/Germany for support of the construction of Advanced LIGO and construction and operation of the GEO600 detector. Additional support for Advanced LIGO was provided by the Australian Research Council. Virgo is funded, through the European Gravitational Observatory (EGO), by the French Centre National de Recherche Scientifique (CNRS), the Italian Istituto Nazionale di Fisica Nucleare (INFN) and the Dutch Nikhef, with contributions by institutions from Belgium, Germany, Greece, Hungary, Ireland, Japan, Monaco, Poland, Portugal, Spain. KAGRA is supported by Ministry of Education, Culture, Sports, Science and Technology (MEXT), Japan Society for the Promotion of Science (JSPS) in Japan; National Research Foundation (NRF) and Ministry of Science and ICT (MSIT) in Korea; Academia Sinica (AS) and National Science and Technology Council (NSTC) in Taiwan.

The authors are grateful for computational resources provided by the LIGO Lab and supported by NSF Grants PHY-0757058 and PHY-0823459. Resources supporting this work were also provided by the NASA High-End Computing (HEC) Program through the NASA Advanced Supercomputing (NAS) Division at Ames Research Center to produce the SPOC data products \citep{jenkins:spoc}.

Some of the results in this paper have been derived using the \texttt{healpy} \citep{2019JOSS....4.1298Z} and
\texttt{HEALPix} \citep{2005ApJ...622..759G} packages.}

This paper carries LIGO document number LIGO-P2200395.

\facilities{LIGO, EGO:Virgo, KAGRA, TESS}

\software{\texttt{astropy} \citep{astropy:2013},
         \texttt{bilby} \citep{Ashton:2018jfp, Romero-Shaw:2020owr},
         \texttt{gwemlightcurves} \citep{Coughlin:2018miv, Coughlin:2018fis, Dietrich:2020efo},
         \texttt{hdbscan} \citep{2017JOSS....2..205M},
         \texttt{healpy} \citep{2019JOSS....4.1298Z},
         \texttt{HEALPix} \citep{2005ApJ...622..759G},
         \texttt{lightkurve} \citep{lightkurve},
         \texttt{ligo.skymap} \citep{Singer:2015ema, Singer:2016eax, Singer:2016erz},
         \texttt{matplotlib} \citep{Hunter:2007},
         \texttt{numpy} \citep{harris2020array},
         \texttt{pandas} \citep{reback2020pandas, mckinney2010data},
         \texttt{scipy} \citep{2020SciPy-NMeth},
         \texttt{tess-point} \citep{2020ascl.soft03001B}
        }
\clearpage
\bibliography{gwtess}{}
\bibliographystyle{aasjournal}

\end{document}